\documentclass[10pt,times,twocolumn]{article}

\usepackage[all]{xy}
\usepackage{color,url}
\usepackage{subfigure}
\usepackage{bm,amsmath}
\usepackage{marvosym}
\usepackage{graphicx}
\usepackage{wrapfig}
\usepackage{epsf}
\usepackage{times,mathptm}
\usepackage{url}

\begin{document}

\title{\Large \bf Electromechanical Wave Green's Function Estimation from Ambient Electrical Grid Frequency Noise}

\author {{\bf Scott Backhaus} $^{(1)}$ and {\bf Yilu Liu} $^{(2)}$\\
Material Physics and Applications Division, LANL, Los Alamos, NM 87545, USA$^{(1)}$\\
EECS, University of Tennessee, Knoxville, TN 37996, USA $^{(2)}$
}

\maketitle

\begin{center}
{\bf \large Abstract}
\end{center}
{\it Many electrical grid transients can be described by the propagation of electromechanical (EM) waves that couple oscillations of power flows over transmission lines and the inertia of synchronous generators.  These EM waves can take several forms: large-scale standing waves forming inter-area modes, localized oscillations of single or multi-machine modes, or traveling waves that spread quasi-circularly from major grid disturbances. The propagation speed and damping of these EM waves are potentially a powerful tool for assessing grid stability, e.g. small signal or rotor angle stability, however, EM wave properties have been mostly extracted from post-event analysis of major grid disturbances.  Using a small set of data from the FNET sensor network, we show how the spatially resolved Green's function for EM wave propagation can be extracted from ambient frequency noise without the need for a major disturbance.  If applied to an entire interconnection, an EM-wave Green's function map will enable a model-independent method of predicting the propagation of grid disturbances and assessing stability.
}

\vspace{0.5cm}

\section{Introduction}
\label{sec:Intro}

The frequency of the electrical grid is determined to a great degree by the rotational speed of large synchronous generators at centralized power stations\cite{94Kun}.  In equilibrium, the mechanical power supplied to the shafts of these generators is in balance with the electrical power withdrawn by the electrical loads and system dissipation.  Fluctuations in this power balance both large and small do occur, and the kinetic energy stored in the rotational inertia of these large generators and associated turbines are the initial buffer against major frequency changes, e.g. if the electrical load were to suddenly increase without a corresponding increase of input mechanical power, the power to supply the load is extracted from the rotational kinetic energy of the generators, causing them to slow and the electrical frequency to decrease.  Such power imbalances are typically not sustained because feedback control systems, i.e. governors and automatic generation control, modulate the mechanical input power and restore balance within a few seconds to tens of seconds.  In this manuscript, we are interested in time scales that overlap with both the ''inertial'' dynamics of the electrical system as well as the primary governor response\cite{94Kun}.

The electrical transmission grid is a distributed system, i.e. the electrical generation and load are point sources spread over a wide geographical area ($\sim 2000\; km$) interconnected by a grid of high voltage ($>115\; kV$) transmission lines.  On longer time scales, the frequency across this distributed system is uniform, and any power imbalance is initially a local phenomenon.  When the electrical load suddenly increases at a particular transmission bus, the phases (and power flows) between that bus and its nearest neighbor buses suddenly increase.  The increased power flow causes the rotational speed of nearby generators to slow as they deliver their stored kinetic energy to the increased load.  This slowing increases the electrical phase angle between the nearby generators and their nearest-neighbor generators.  The transfer of power from the nearest neighbors supports the original generators, but their rotational speed also slows (to a lesser degree as there are more of them) building up phase difference and power flow from more distant generators.  The effect is an approximately circular, outward propagating electromechanical (EM) traveling wave whose speed (~1000 km/sec) is determined to first order by the rotational inertia of the generators and the susceptance of the transmission lines\cite{Thorp1998}.  These properties can show significant spatial variation over an electrical grid, particularly in situations where generators and loads are clustered yet separated from one another.  There can also be temporal variation on many time scales as patterns of generation and load change in time.  In addition, the wave propagation may appear to be one or two-dimensional depending on the topology of the electrical grid\cite{Seyler2004}.

Many transient grid phenomena can be understood from the standpoint of EM waves, e.g, inter-area oscillations (i.e. global modes) are a manifestation of an interconnection-scale standing EM wave.  From typical EM wave speeds and interconnection size, we estimate the frequency of the fundamental, one-wavelength, north-south mode in WECC to be approximately $f_{N-S}\sim [1000\; km/sec]/[3000\; km]\sim 0.3\; Hz$, consistent with observed values\cite{94Kun}.  The WECC modes with shorter wavelengths have higher frequency\cite{94Kun}, which is also consistent with the interpretation in terms of standing EM waves.  The physics of local plant modes is the same as inter-area oscillations\cite{94Kun}, however, local plant modes typically only extend over a single or a few generators and a small part of the transmission system, invalidating a formal continuum description in terms of EM waves.  In spite of this, the technique we develop in this manuscript should also be applicable to characterizing local plant modes.

Inter-area and local plant modes are small signal oscillations that, under the right conditions, may become unstable and spontaneously grow in amplitude.  Even if stable, these modes can be excited to significant amplitude by a major system disturbance.  In either case, the resulting oscillations can lead to protective relaying actions that may cause loss of load, system separation, or large-scale blackouts\cite{94Kun,Kundur1991}. A good understanding of EM-wave propagation properties will create a better understanding of the oscillatory modes and transients of an electrical interconnection.  Even better, {\it a real-time wide-area measurement system (WAMS) that extracts the EM-wave properties would provide a model-independent method for estimating oscillatory mode shapes, frequencies, and damping and a method for predicting how a major disturbance will impact the rest of the interconnection.}  The intent of this manuscript is to demonstrate the feasibility of extracting the Green's function\cite{morseandfeshbach} for EM wave propagation from ambient frequency noise, a crucial first step towards realizing such a real-time monitoring tool.

\section{Model-based stability}
\label{sec:Model}
Rigorous one and two dimensional continuum models of transmission and generation \cite{Thorp1998,Seyler2004} have been developed that describe EM waves, and they predict propagation characteristics similar to that described above.  These models are useful for developing qualitative understanding of many transient grid phenomena, however, the use of these reduced models in grid operations and monitoring is in doubt because the quantitative accuracy of these models is questionable due to the complexity of real electrical loads and control systems and the presence of additional equipment not included in the models, e.g. DC lines/ties, phase shifting transformers, and other FACTS devices.

Specialized techniques have been developed for computing the mode frequencies and damping for detailed models of large interconnections with many generators and transmission lines\cite{94Kun}.  For a given system condition, these techniques can accurately characterize the interconnection's oscillatory modes.  Also, detailed time-harmonic transient models of an interconnection can capture the details of EM waves, i.e. a frequency disturbance propagating outward from the initiating event\cite{Tsai2007} even including the reflections of the waves off the grid boundaries.  We note that the accuracy of both types of off-line studies is dependent on the time-consuming task of creating models with accurate parameters for many tens of thousands pieces of grid equipment, however, the uncertainty in model parameters due to continually changing grid operating conditions and equipment availability calls into question the viability of real-time assessment of interconnection stability using these model-based techniques.  Here, we explore the feasibility of estimating EM-wave Green's functions from ambient frequency noise in WAMS measurements.  These Green's functions will form the basis of a future model-independent method for assessing small signal and transient stability.

\section{Existing WAMS measurements}
\label{sec:wams}

The advent WAMS\cite{Liu2009,naspi} allows for detection and monitoring of EM waves in much more detail than in the past.  WAMS-based study of EM waves has focused on two major areas: the propagation and spatial localization of major frequency disturbances\cite{Liu2007,Liu2008,Liu2009,Tsai2007} and the damping and mode structure of major inter-area oscillations\cite{Hauer2010,Hauer2007,Trud2008}.  In the first example, a sudden loss several hundred MWs of generation due to a system disruption results in a frequency decline of ~0.1 Hz, which is easily detectable by WAMS\cite{Liu2008}.  Measuring the differences in arrival times at different points in the WAMS and assuming or estimating EM wave propagation speed allows for approximate location of the initial disturbance\cite{Liu2007}.  Alternatively, the WAMS transient data could be used to estimate EM wave speeds between the known disturbance location and the detection point, however, major frequency disturbances only occur once every few days and the conditions of the electrical grid are continually changing, which makes the applicability of WAMS-extracted EM wave properties uncertain for grid conditions as little as an hour later.  In the future, this time scale will likely shorten to a few minutes  as time-intermittent renewable generation, such as wind and solar photovoltaic, will create fast and stochastic changes in grid conditions. It is exactly under these highly variable conditions that real-time monitoring of grid conditions via EM waves will be the most valuable.

In the second example, correlations in the frequency spectra of noise-driven (or probe-signal driven) frequency oscillations are used to identify the amplitude and time-phase of standing EM-wave oscillations at system buses participating in a small number of inter-area oscillations\cite{Trud2008}.  We note that in \cite{Trud2008}, the frequency of the single mode that was studied was previously identified by an eigenvalue analysis.  The reliance on correlations of spectra between buses appears to restrict this analysis to a small numbers of modes that show a significant level correlation across the interconnection.  Modes that are spatially localized will not show such long-range correlations and would be difficult to identify.  By only studying a single or a few modes, this analysis ignores a wealth of information in the remainder of the spectra that describes how {\it transients} propagate throughout the system.

\section{Green's function estimation}
\label{sec:green}
As an EM-wave propagates from a bus at location $\mathbf{x}_1$ to a bus at location $\mathbf{x}_2$, it effectively encodes all of the grid properties between $\mathbf{x}_1$ and $\mathbf{x}_2$ making EM waves a potentially powerful grid diagnostic.  For a Dirac delta-function frequency disturbance at $\mathbf{x}_2$ at $t=0$, i.e. $\Delta f(\mathbf{x},t)=\Delta f_0 \delta(\mathbf{x}-\mathbf{x}_2)\delta(t)$, the time response of the frequency at $\mathbf{x}_1$ is called the Green's function $G_{\Delta f}(\mathbf{x}_1,\mathbf{x}_2,t)$.  In this manuscript, we demonstrate the feasibility of estimating $G_{\Delta f}(\mathbf{x}_1,\mathbf{x}_2,t)$ for EM wave propagation in near real time from ambient frequency noise.  We note that the frequency disturbance used to define $G_{\Delta f}(\mathbf{x}_1,\mathbf{x}_2,t)$ also corresponds to a step change in phase at $\mathbf{x}_2$, i.e. $\Delta \Phi(\mathbf{x},t)=\Delta \Phi_0 \delta(\mathbf{x}-\mathbf{x}_2)\Theta(t)$, which corresponds to an abrupt change in the local generation-load balance as described above and investigated in \cite{Liu2007,Liu2008,Liu2009,Tsai2007}.

Our Green's function estimation approach is adapted from and closely follows recent theoretical work by Snieder\cite{Snieder2004} on passive imaging via correlation of earthquake coda waves in locally isotropic media and related experimental work by Derode {\it et al}\cite{Derode2003}.  In this manuscript, we review the relevant parts of Snieder's work and point out the essential differences with the present problem. Additional references on the foundations of this method can be found in the citations of \cite{Snieder2004} and \cite{Derode2003}.

Snieder considers estimation of the Green's function for wave propagation between two observation points $\mathbf{x}_1$ and $\mathbf{x}_2$ in a two-dimensional wave propagation medium that is embedded with many randomly placed scatterers (indexed by s). Snieder assumes that a wave is launched into the medium by some external source or event, and the scatterers act as secondary sources of singly and multiply scattered waves such that each scatterer re-emits a wave given by $S_s(t)$.  The $S_s(t)$ are uncorrelated because the scattering sites are randomly placed.  These waves propagate from their scattering (source) locations to $\mathbf{x}_1$ and $\mathbf{x}_2$ (with their individual time delays), and the linear superposition of these waves form the aggregate signals received at $\mathbf{x}_1$ and $\mathbf{x}_2$, i.e. $p_{\mathbf{x}_1}(t)$ and $p_{\mathbf{x}_2}(t)$.  Next, we form the cross correlation of the two received signals over a time window $T$
\begin{equation}\label{eq:C}
C(\mathbf{x}_1,\mathbf{x}_2,t) \equiv \int_0^T p_{\mathbf{x}_2}(\tau+t)p_{\mathbf{x}_1}(\tau)d\tau.
\end{equation}
	
In principle, $C(\mathbf{x}_1,\mathbf{x}_2,t)$ involves a double sum over scatters $s$ for $p_{\mathbf{x}_1}(t)$ and $s^\prime$ for $p_{\mathbf{x}_2}(t)$, however, Snieder\cite{Snieder2004} shows that if the individual scattered waves (i.e. sources) $S_s(t)$ {\it do not have a time average component}, the cross terms with $s\neq s^\prime$ average to zero after a sufficiently long $T$, leaving only the diagonal terms with $s=s^\prime$. This requirement on the $S_s(t)$ will be crucial in determining the final form of $G_{\Delta f}(\mathbf{x}_1,\mathbf{x}_2,t)$.

With only the diagonal terms ($s=s^\prime$) remaining, $C(\mathbf{x}_1,\mathbf{x}_2,t)$ is simply a sum of autocorrelations of the individual source signals $S_s(t)$ where the time delay in the autocorrelation is a sum of the original delay $t$ in the cross correlation from Eq.~\ref{eq:C} and the difference in the arrival times at $\mathbf{x}_1$ and $\mathbf{x}_2$ from scatterer $s$.  The fact that the $S_s(t)$ do not have a DC component means that they are oscillatory in nature, therefore, the individual autocorrelations will be oscillatory are well.  The phase of these oscillations depends strongly on the autocorrelation time delay, which itself depends on the difference in arrival times at $\mathbf{x}_1$ and $\mathbf{x}_2$ from scatterer $s$.  For source locations away from a straight line passing through $\mathbf{x}_1$ and $\mathbf{x}_2$ (the ``receiver line''), this difference in arrival time varies rapidly with source location causing fast variations in the phase of the autocorrelations such that contributions from these source locations interfere destructively.  However, for source locations on the receiver line and not between $\mathbf{x}_1$ and $\mathbf{x}_2$, the difference in arrival times is nearly constant regardless of the absolute location of $s$.  The relative phase of these individual autocorrelations is stationary, and the signals from these source locations interfere constructively.  The $S_s(t)$ emanating from locations on or very near the receiver line make the vast contribution to the cross correlation $C(\mathbf{x}_1,\mathbf{x}_2,t)$.

From the physical, stationary phase argument above, one can already see that $C(\mathbf{x}_1,\mathbf{x}_2,t)$ will contain information about $G(\mathbf{x}_1,\mathbf{x}_2,t)$.  Consider a source point $s$ on the receiver line very near $\mathbf{x}_2$  but on the opposite side of $\mathbf{x}_2$ relative to $\mathbf{x}_1$.  Viewed from $\mathbf{x}_1$, we could hardly distinguish whether a delta-function pulse from $s$ originated from $s$ or from $\mathbf{x}_2$.  The signal detected at $\mathbf{x}_1$ would be nearly equal to that caused by  delta-function pulse from $\mathbf{x}_2$, which is exactly $G(\mathbf{x}_1,\mathbf{x}_2,t)$. A similar argument can be made for the contributions from the other source terms on or near the receiver line but not necessarily close to $\mathbf{x}_2$.

First applying a Fourier transform, Snieder\cite{Snieder2004} works out the details of the stationary phase calculation by converting the sum over individual source terms $s$ to an integral over a source area density $n(x)$.  The result, given by his Eq. 24 and adapted to our setting, is
\begin{eqnarray}
C(\mathbf{x}_1,\mathbf{x}_2,\omega)&=&\frac{\pi c}{i\omega}\overline{|S(\omega)|^2}[G(\mathbf{x}_2,\mathbf{x}_1,\omega)\int_{-\infty}^0 n(x) dx\nonumber\\
&+& G(\mathbf{x}_1,\mathbf{x}_2,\omega)\int_{R}^\infty n(x) dx ]\label{eq:Cw}.
\end{eqnarray}
where $C(\mathbf{x}_1,\mathbf{x}_2,\omega)$ and $G(\mathbf{x}_1,\mathbf{x}_2,\omega)$ are the Fourier transforms of the cross correlation and Green's function, $\overline{|S(\omega)|^2}$ is power spectral density averaged over the sources $S_s(t)$, and $c$ is the phase speed of  EM-wave between $\mathbf{x}_1$ and $\mathbf{x}_2$.  The integrals in Eq.~\ref{eq:Cw} are carried out along two sections of the receiver line with $\mathbf{x}_1$ at $x=0$ and $\mathbf{x}_2$ at $x=R$.  Snieder\cite{Snieder2004} performed his derivation for surface waves on a solid, and he accounted for different polarizations and different wave modes.  EM waves are also two-dimensional waves, however, they are only scalar waves with a single mode.  Therefore, in Eq.~\ref{eq:Cw}, we have dropped the polarization, mode indexes, and complex conjugation in Snieder's Eq. 24.  Equation~\ref{eq:Cw} is the starting point for our estimation of EM-wave Green's function from ambient frequency noise on an electrical interconnection.

\section{EM-wave Green's function estimation}
\label{sec:EM-wave}

The electrical grid has been approximated as a two-dimensional continuum of transmission and generation, and frequency disturbances have been  shown to propagate as EM waves as described above\cite{Seyler2004}.  To use the Green's function estimation technique as described by Snieder\cite{Snieder2004}, we must be careful in its application.  In the continuum description of the grid, there are not embedded scatterers as described by Snieder\cite{Snieder2004}, however, his scatterers simply acted as sources of uncorrelated injections of signals $S_s(t)$ from randomly placed locations.  In the case of the electrical grid, random fluctuations of load at substation buses will play the same role.

On short time scales, changes in load $P$ at a bus in the transmission system are abrupt, and an individual change can be modeled as a step function in time.  Previously, we argued that such changes in load create instantaneous frequency deviations at this bus (located at $\mathbf{x}_s$) in the form of a Dirac delta-function in time, i.e.
$\Delta f(\mathbf{x},t)=\Delta f_s \delta(\mathbf{x}-\mathbf{x}_s)\delta(t)$.  If we used the cross correlation in Eq.~\ref{eq:C} to estimate $G_{\Delta f}(\mathbf{x}_1,\mathbf{x}_2,t)$, the source terms $S_s(t)\sim \Delta f_s \delta(\mathbf{x}-\mathbf{x}_s)\delta(t)$ have a time-average component violating a basic assumption made by Snieder\cite{Snieder2004}.  To circumvent this difficulty, we choose instead to work with the time derivative of the frequency deviation $\Delta f^\prime$ and estimate the Green's function $G_{\Delta f^\prime}(\mathbf{x}_1,\mathbf{x}_2,t)$.  The source terms are $S_s(t)\sim \Delta f_s \delta(\mathbf{x}-\mathbf{x}_s)d\delta(t)/dt$, which have no time-average component.  The average of the power spectra in Eq.~\ref{eq:Cw} is then $\overline{|S(\omega)|^2}\propto \omega^2$. Substituting into Eq.~\ref{eq:Cw}, we find
\begin{eqnarray}
\frac{C_{\Delta f^\prime}(\mathbf{x}_1,\mathbf{x}_2,\omega)}{i\omega}&=&-\pi c[G_{\Delta f^\prime}(\mathbf{x}_2,\mathbf{x}_1,\omega)A_-\nonumber\\
&+& G_{\Delta f^\prime}(\mathbf{x}_1,\mathbf{x}_2,\omega)A_+ ]\label{eq:Cw2}.
\end{eqnarray}
Taking the inverse Fourier transform of Eq.~\ref{eq:Cw2} yields
\begin{eqnarray}
\int C_{\Delta f^\prime}(\mathbf{x}_1,\mathbf{x}_2,t)dt&=&-\pi c[G_{\Delta f^\prime}(\mathbf{x}_2,\mathbf{x}_1,t)A_-\nonumber\\
&+& G_{\Delta f^\prime}(\mathbf{x}_1,\mathbf{x}_2,t)A_+ ],\label{eq:Ct}
\end{eqnarray}
where $A_-$ and $A_+$ are the integrals over the source density in Eq.~\ref{eq:Cw} from $-\infty \rightarrow 0$ and from $R\rightarrow \infty$, respectively. If the electrical grid is in a quasi-steady state over the time window $T$, $\Delta f$ and $\Delta f^\prime$ satisfy the same homogenous partial differential equation and $G_{\Delta f^\prime}(\mathbf{x}_1,\mathbf{x}_2,t)=G_{\Delta f}(\mathbf{x}_1,\mathbf{x}_2,t)$.  Therefore, the time integral of the cross correlation $C_{\Delta f^\prime}(\mathbf{x}_1,\mathbf{x}_2,t)$ in Eq.~\ref{eq:Ct} also yields the Green's function for $\Delta f$, a quantity of fundamental importance.

If we had a map of $G_{\Delta f}(\mathbf{x}_1,\mathbf{x}_2,t)$ for many pairs of locations throughout the interconnection, we could quickly compute the time dependence of the frequency deviation following a major disturbance to the grid's balance of generation and load.  In addition, $G_{\Delta f}(\mathbf{x}_1,\mathbf{x}_2,t)$ can also be used to estimate the local EM-wave speed and attenuation throughout the interconnection which can then be used to quickly compute the frequencies and mode shapes of many inter-area oscillations and the full transient response to a grid disturbance.

\section{Source data and analysis}
\label{sec:data}

To test the concepts described above, we have analyzed a small sample of data from the FNET WAMS\cite{Liu2009}.  The FNET sensors provide time-synchronized frequency measurements  every 0.1 secs at many transmission buses throughout the U.S.  Here, we focus on three closely-spaced buses in the Eastern interconnection shown in Fig. 1.  The frequency measurements are actually performed at a location on the distribution system attached to the transmission bus, however, it has been shown that frequency disturbances travel through the distribution system quickly compared to the transmission system making the FNET frequency measurements accurate representations of the bus-local frequency on the transmission bus\cite{Liu2009}.
\begin{figure}[t]
\centering
\includegraphics[width=0.5\textwidth]{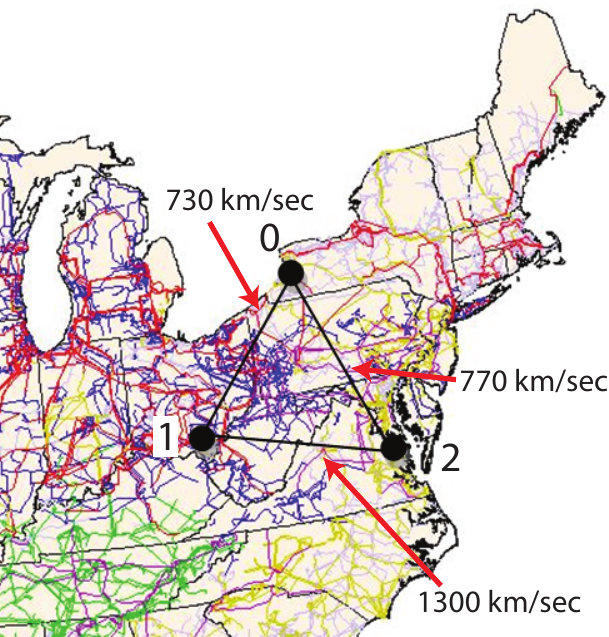}
\caption{The approximate location of the FNET sensors used in this analysis are indicated by the filled circles.  The black lines are the receiver lines for each pair of buses.  The EM-wave speed for the receiver lines are computed from the arrival times estimated from the Green's function $G_{\Delta f}(\mathbf{x}_1,\mathbf{x}_2,t)$ for each pair of buses.}
\label{fig:map}
\end{figure}

\begin{figure}[]
\centering
\includegraphics[width=0.5\textwidth]{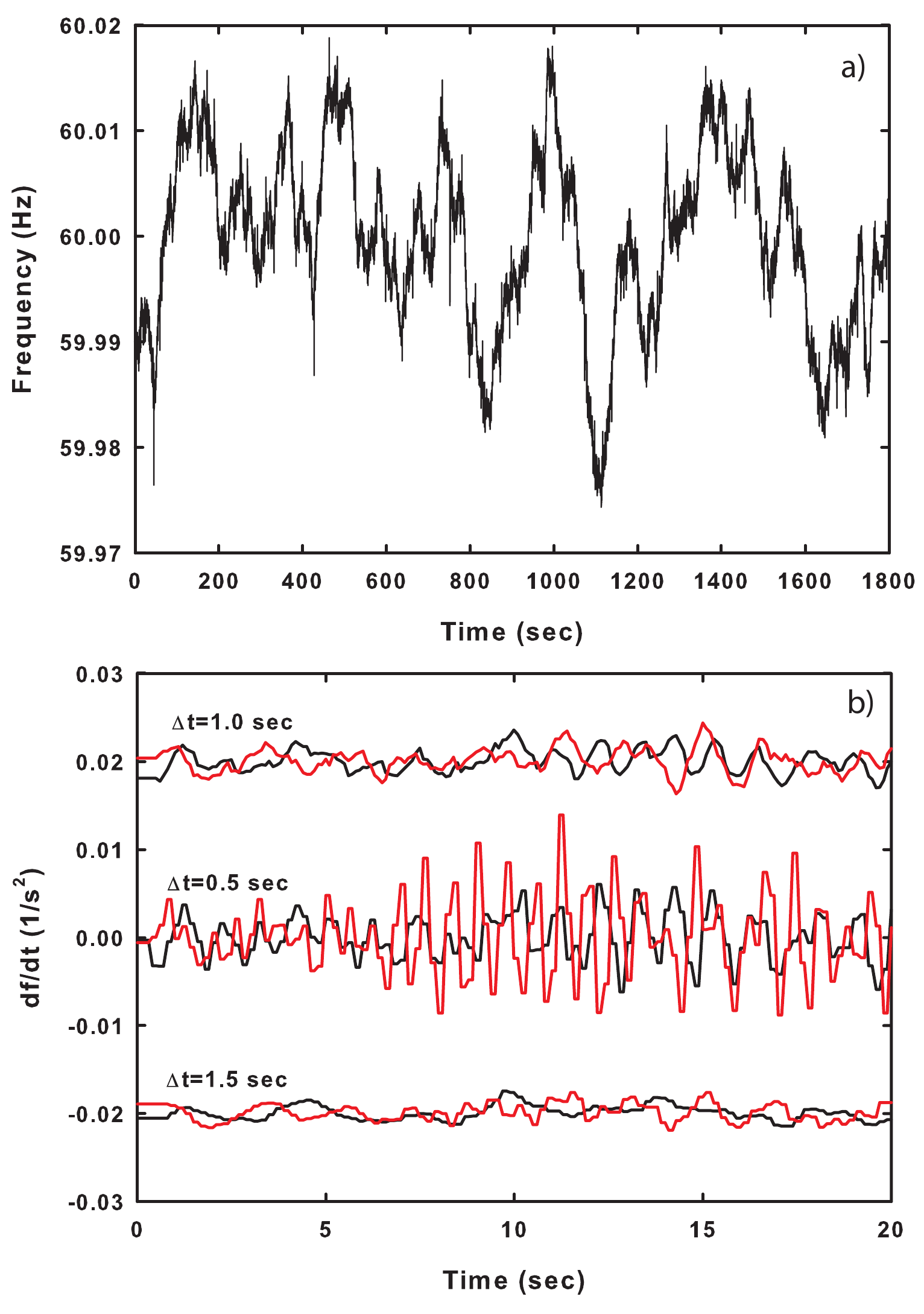}
\caption{a) A $30$-minute window of frequency data from the FNET sensor near bus $0$.  The data shows normal variability about the base frequency with no significant sudden increase or decrease in frequency.  The downward spikes near $50$ and $400\;sec$ are likely just spurious, non-Gaussian sensor noise.  The frequency data from buses $1$ and $2$ are indistinguishable from bus $0$ on this time scale.  b) $df/dt$ for bus $0$ (black) and $1$ (red) for the first $20\;secs$ of the window in a).  $df/dt$ vs $t$ is calculated from the linear coefficient a quadratic fit to the data in a) using different length time windows $\Delta t$ centered on $t$.}
\label{fig:f-dfdt}
\end{figure}

Figure~\ref{fig:f-dfdt}a shows a 30-minute segment of frequency data from bus $0$ in Fig. 1.  On this time scale, the data from the other two buses closely track bus $0$ and are indistinguishable.  The data show normal variations about the base frequency, and important for our discussion, the data does not show any significant transients that can be attributed to a system  disturbance.  The data sets do show spurious spikes which we attribute to sensor noise.  We have not made any specific effort to remove such noise.

To apply Eq.~\ref{eq:Ct}, we must cross correlate $d\Delta f/dt$ at the three different sites.  We compute the derivative at time $t$ from the linear coefficient of a parabolic fit to a window of data $\Delta t$ wide and centered on $t$.  The window $\Delta t$ is varied in the following analysis to investigate the sensitivity of the results to this implicit filtering.  Figure~\ref{fig:f-dfdt}b shows the resulting $d\Delta f/dt$ for buses $0$ and $1$ for $\Delta t$ ranging from $0.5\; secs$ to $1.5\;secs$.  We will use these two buses to investigate the how quickly the Green's function estimation technique converges.

\begin{figure}[t]
\centering
\includegraphics[width=0.5\textwidth]{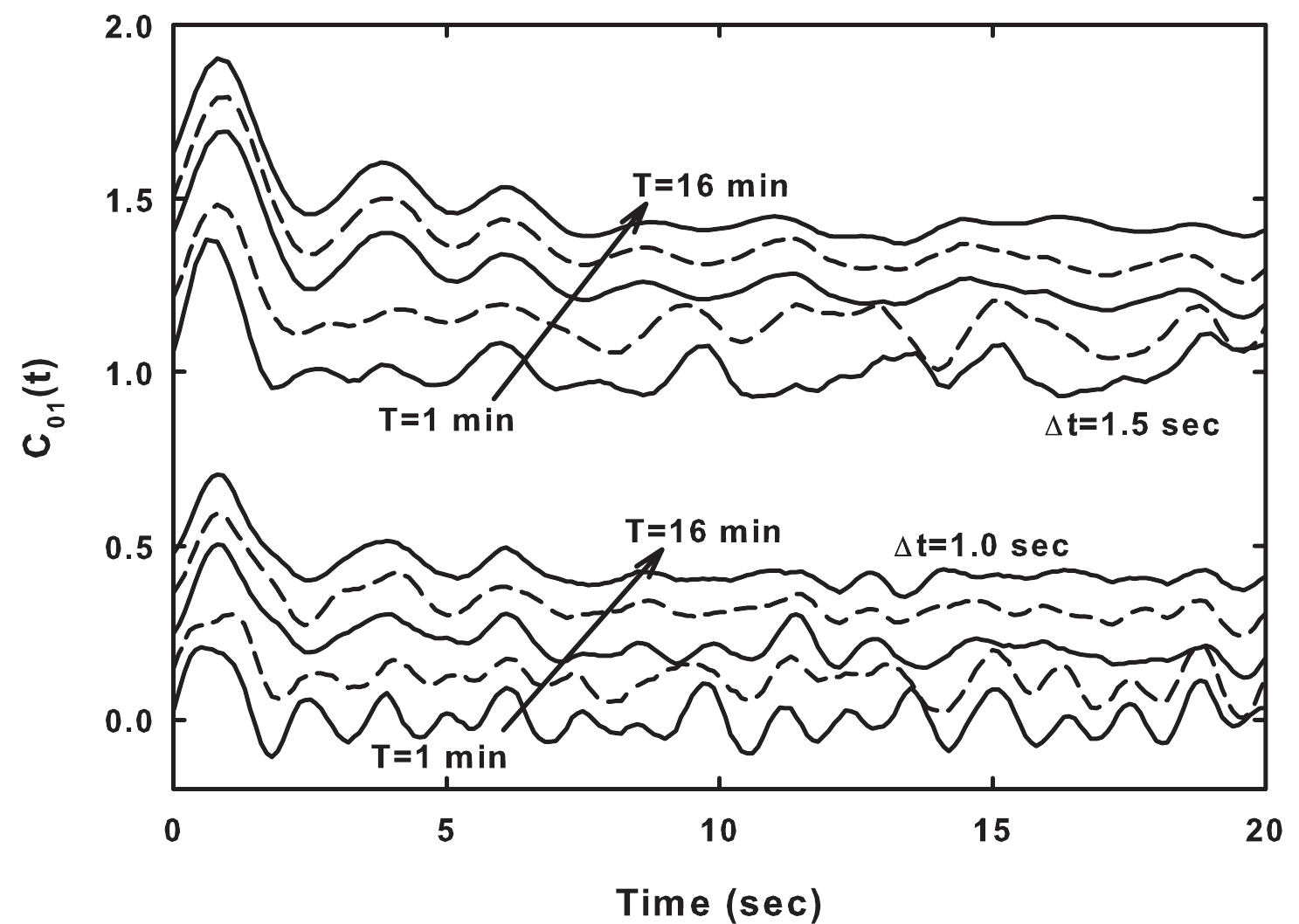}
\caption{The cross correlation of $df/dt$ at buses $0$ and $1$, $C_{01}$.  The two groups of curves are computed using $df/dt$ data generated using $\Delta t$ fitting windows $1.0$ and $1.5\;sec$ wide.  Within each group, averaging times $T$ of $1$, $2$, $4$, $8$, and $16\; minutes$ are used.  The reduction in fluctuations of $C_{01}$ beyond about $10\;sec$ is used as an indication of the quality of destructive interference of the EM waves launched by sources away from the receiver line for buses $0$ and $1$.  These fluctuations become small and the functional form of $C_{01}$ becomes relatively constant at $\Delta t=1.0\;sec$ and $T=8\;minutes$ or $\Delta t=1.5\;sec$ and $T=4\;minutes$.  The zeroes for the $\Delta t=1.0\;sec$ data are incrementally offset by 0.1 along the vertical axis for clarity of the figure.  The zeroes for $\Delta t=1.5\;sec$ data are offset in a similar way with the zero for the $T=1\; minute$ data at 1.0.}
\label{fig:c01-T}
\end{figure}

The cross correlation in Eq.~\ref{eq:Ct} can be computed over a range of averaging times $T$.  As $T$ becomes longer, the cross correlation time integral averages over a greater number of load fluctuations and the destructive interference between the $S_s(t)$ away from receiver lines becomes more complete\cite{Snieder2004}, improving the estimation of $G_{\Delta f}(\mathbf{x}_1,\mathbf{x}_2,t)$.  However, a long $T$ would limit our ability to detect rapid changes in grid conditions in real time.  Figure~\ref{fig:c01-T} shows the evolution in $C_{\Delta f^\prime}(\mathbf{x}_1,\mathbf{x}_2,t)$ for $T$ ranging from $1\;minute$ to $16\;minutes$ and for a $\Delta t$ of $1.0\;sec$ and $1.5\;sec$ for buses $0$ and $1$.  In Figure~\ref{fig:c01-T}, $T$ increases by a factor of two for each of the five curves in the groups of $\Delta t$.

Many FNET observations of the of the Eastern interconnection following a major disturbance show that the post-disturbance frequency is relatively constant for times greater than $10\;sec$ after the disturbance\cite{Liu2007,Liu2008,Liu2009}.  Since the cross correlation in Fig.~\ref{fig:c01-T} is the derivative of this response (i.e. of the Green's function), the cross correlation should approach a steady value of zero beyond about $10\;sec$. For $\Delta t=1.5\;sec$, the fluctuations in $C_{01}$ beyond $10\;sec$ are greatly diminished for $T\geq 4\;min$ and $C_{01}$ appears to reach a relatively constant functional form.  For $\Delta t=1.0\;sec$, $T\geq 8\;min$ is required to achieve similar behavior. These minimum values of $T$ necessary to achieve adequate destructive interference of the noise sources off of the receiver line should be compared to the $\sim 20\;min$ averaging time to extract accurate results for bus-specific amplitudes of a single inter-area mode in \cite{Trud2008}.  In principle, our method should require less averaging time because we are using all of the information in the cross correlation below a frequency of $\sim 1/\Delta t$ as opposed to the single frequency in \cite{Trud2008}.

From the definition of the cross correlation in Eq.~\ref{eq:C} and the interpretation in terms of Green's functions, the cross correlation in Fig.~\ref{fig:c01-T} should non-zero for both positive and negative times shifts $t$, however, it should be zero at $t=0$ as a disturbance does not propagate with infinite speed.  In Fig.~\ref{fig:c01-T}, we only show the cross correlation for one half of the time axis, and it does not go to zero at $t=0$.  Both of these points can be explained by the distribution of sources around the receiver lines in Fig.~\ref{fig:map} and the filtering we perform on $df/dt$.

When a pulse is emitted at $\mathbf{x}_1$, it propagates at finite speed and generates a response at $\mathbf{x}_2$ some time later, therefore, EM-wave propagation from $\mathbf{x}_1 \rightarrow \mathbf{x}_2$ generates a non-zero cross correlation in Eq.~\ref{eq:C} for $t>0$ and is due to sources contained in the integral $A_-$ in Eq.~\ref{eq:Ct}.  A similar argument concludes that EM-wave propagation from $\mathbf{x}_2 \rightarrow \mathbf{x}_1$ generates a non-zero cross correlation for $t<0$ and is due to sources contained in the integral $A_+$.  From these arguments, it is clear that $G_{\Delta f^\prime}(\mathbf{x}_2,\mathbf{x}_1,t)$ is the causal Green's function for EM-wave propagation from $\mathbf{x}_1 \rightarrow \mathbf{x}_2$ and is non-zero for $t>0$, and $G_{\Delta f^\prime}(\mathbf{x}_1,\mathbf{x}_2,t)$ is the anti-causal Green's function and is non-zero for $t<0$.  For time-reversal invariant systems, $G_{\Delta f^\prime}(\mathbf{x}_2,\mathbf{x}_1,t)=G_{\Delta f^\prime}(\mathbf{x}_1,\mathbf{x}_2,-t)$, however, the cross correlation in Eq.~\ref{eq:Ct} need not be symmetric because the location and orientation of the receiver line in the network may yield weights $A_-$ and $A_+$ that are not the same.  Such a situation arises when one end of the receiver is near the edge of the network and proximity of the edge restricts either $A_-$ or $A_+$.

\begin{figure}[t]
\centering
\includegraphics[width=0.5\textwidth]{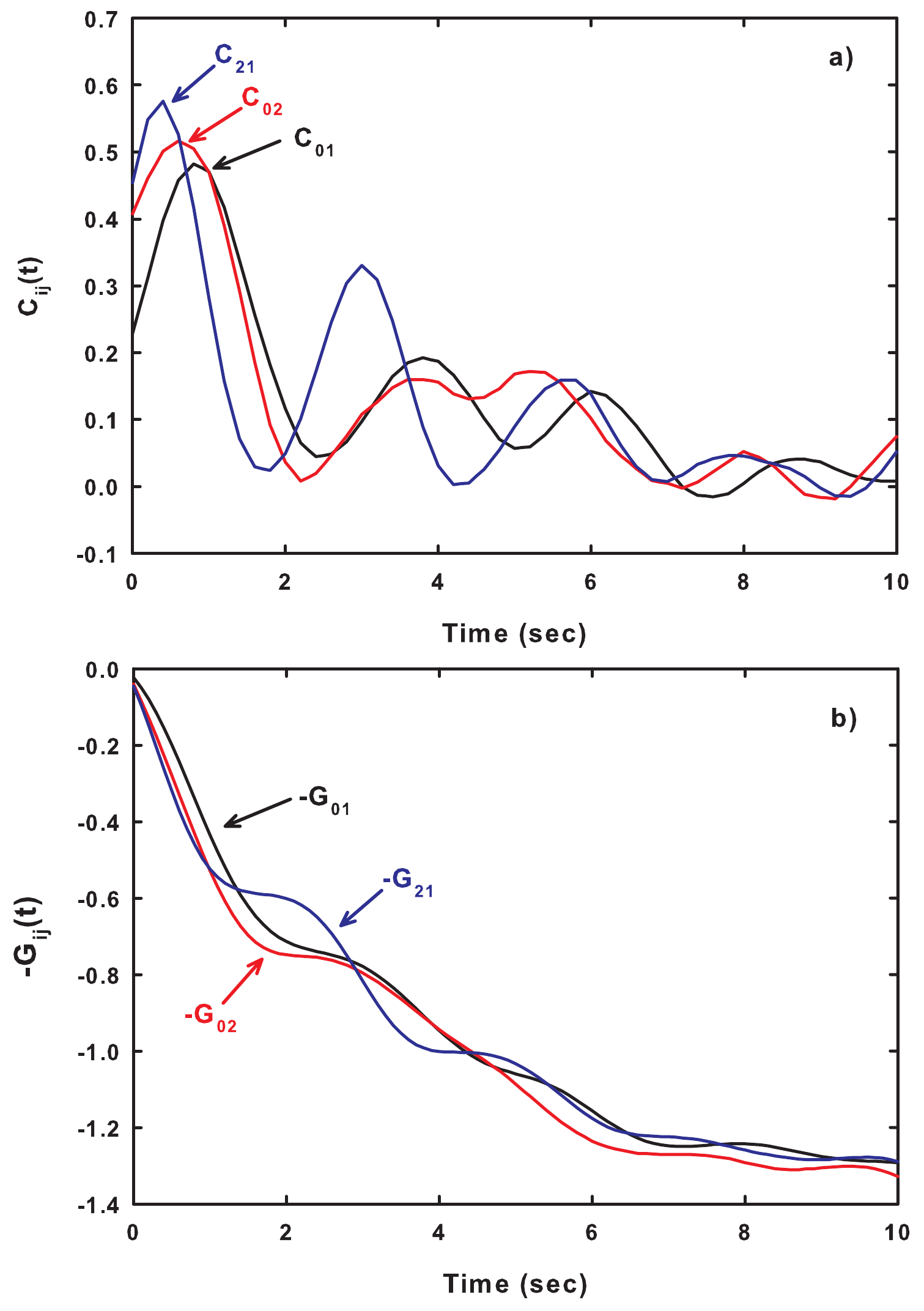}
\caption{a) The cross correlations for each pair of observation points in Fig.~\ref{fig:map} computed using $\Delta t=1.5\;sec$ and $T=4\;mintues$.  b) The negative of the integral of the cross correlations in a), which are proportional to $-G_{\Delta f}(\mathbf{x}_1,\mathbf{x}_2,t)$, i.e. the frequency response at $\mathbf{x}_1$ due to a sudden loss of generation at $\mathbf{x}_2$.
}
\label{fig:CandG}
\end{figure}

For the three pairs of observation points in Fig.~\ref{fig:map}, one end of the receiver line is restricted in each case, i.e. near $0$ for $0\rightarrow 1$ and $0\rightarrow 2$ and near $2$ for $1\rightarrow 2$.  For $0\rightarrow 1$ and $0\rightarrow 2$, we use the cross correlation for $t\leq 0$ and the anti-causal Green's function.  For $1\rightarrow 2$, we use $t\geq 0$ and the causal Green's function. These choices corresponds to waves emanating from noise source on the unrestricted end of the receiver line.  It is these cross correlations (and resulting Green's functions) that are presented in Figs.~\ref{fig:c01-T} and \ref{fig:CandG}.  For waves from noise sources on the restricted end, we find cross correlations that are smaller than and lack the coherent oscillations of those in Figs.~\ref{fig:c01-T} and \ref{fig:CandG}.

The approximate symmetry of the cross correlation about $t=0$ also explains why the curves in Fig.~\ref{fig:c01-T} and \ref{fig:CandG} do not approach zero at $t=0$.  To obtain a reasonable signal-to-noise ratio, we filter $df/dt$ using parabolic fits centered on $t$.  The effect is to spread out any rapid time variation of $df/dt$ causing sharp pulses to appear to arrive before causality would allow.  Since our filtering window $\Delta t$ is of the same order as the propagation time delay, a pulse is that is in reality sharp generates a contribution at $t=0$ when first passed through our filter.

With a better understanding of the techniques for estimating the Green's function for EM wave propagation over an electrical grid, we compute the cross correlations for each pair of observation points using $\Delta t=1.5\;sec$ and $T=4\;min$.  The cross correlations are shown in Fig.~\ref{fig:CandG}a and the integrals, i.e. the Green's function $G_{\Delta f}(\mathbf{x}_1,\mathbf{x}_2,t)$ for the local frequency, are shown in Figs.~\ref{fig:CandG}a and b.  The absolute scaling of $C(\mathbf{x}_1,\mathbf{x}_2,t)$ and $G_{\Delta f}(\mathbf{x}_1,\mathbf{x}_2,t)$ is arbitrary because we currently have no method to estimate $A_+$ or $A_-$ in Eq.~\ref{eq:Ct}.  In an attempt to maintain the relative scaling between the three different observation pairs, we have used the product of root-mean-squares of $df/dt$ at $\mathbf{x}_1$ and $\mathbf{x}_2$ as estimates of $A_+$ and $A_-$

As discussed earlier, $G_{\Delta f}(\mathbf{x}_1,\mathbf{x}_2,t)$ is the frequency response at $\mathbf{x}_1$ after a Dirac delta-function perturbation to the frequency at $\mathbf{x}_2$, which is equivalent to a sudden increase (decrease) in generation (load) at $\mathbf{x}_2$.  Figure~\ref{fig:CandG}b actually presents $-G_{\Delta f}(\mathbf{x}_1,\mathbf{x}_2,t)$ which is the response to a sudden loss of generation at $\mathbf{x}_2$.  The general shape and time extent of the transient closely resembles actual measurements (made with FNET) of such frequency declines after a loss of generation\cite{Liu2007,Liu2008,Liu2009}, however, {\it the response in Fig.~\ref{fig:CandG}b was estimated from ambient frequency noise.}  This noise can be analyzed in real time to provide an on-line prediction of the local and system-wide impact of such a major disturbance.

The arrival time of an EM wave following a major disturbance has been estimated by finding the time of fastest frequency decline\cite{Liu2008}.  For our analysis, this is given by the time of the peak value of $C(\mathbf{x}_1,\mathbf{x}_2,t)$.   These arrival times are $t_{0\rightarrow 1}=0.8\; sec$, $t_{0\rightarrow 2}=0.6\; sec$, and $t_{1\rightarrow 2}=0.4\; sec$.  The straight-line geographic distances between these buses are $L_{0\rightarrow 1}=585\; km$, $L_{0\rightarrow 2}=460\; km$, and $L_{1\rightarrow 2}=510\; km$.  The EM-wave speeds estimated from these values are shown in Fig.~\ref{fig:map}.  The slower EM-wave speeds appear to be correlated with regions of dense transmission and generation which is qualitatively consistent with the interpretation in \cite{Seyler2004}.  However, these analysis should be repeated for time windows $T$ immediately following a major system disturbance, such as those in \cite{Liu2008}, so that a quantitative comparison of the full $G_{\Delta f}(\mathbf{x}_1,\mathbf{x}_2,t)$ can be made.

\section{Conclusions and future work}
\label{sec:con}

Using signal processing techniques developed for passive imaging in seismology\cite{Snieder2004}, we have demonstrated how the transient response of the electrical grid to sudden changes in load or generation can be estimated from ambient frequency noise gathered over a WAMS.  Our method provides the entire time response, not just the phases and relative amplitudes for an individual EM standing-wave mode.  We have shown how the quality of the Green's function for the transient response varies versus different levels of filtering of the raw signal and different cross-correlation integration times.  In this initial work, raw signal filtering with a time constant of about $1-1.5\;sec$ is required to achieve an integration time of less than about $5-10\;mintues$.  Shorter cross-correlation integration times are preferred as they allow for closer to real-time monitoring.

There is much future work to do in this new area.  Here, we mention a few possible directions:
\begin{itemize}
\item PMU data has faster time resolution and different signal-to-noise characteristics.  This method should be applied to PMU data to investigate if improvements in Green's function estimation are possible
\item The estimated Green's function should be compared to the transient responses generated by major disturbances such as loss of hundreds of MW of generation.  The Green's function should be estimated from the ambient frequency noise both before and after the event and compared with the frequency decline during the event.
\item Using data from the FNET WAMS or PMU data, all nearest-neighbor Green's functions should be estimated and an EM-wave propagation speed and attenuation map developed for an entire interconnect.  The mode shapes, frequencies, and damping of all inter-area modes should be estimated and compared with off-line simulations.
\end{itemize}

\section{Acknowledgments}

We are thankful to the participants of the "Optimization and
Control for Smart Grids" LARD DR project at Los Alamos
and Smart Grid Seminar Series at CNLS/LANL for multiple
fruitful discussions. The work at LANL was carried out under the auspices of the National
Nuclear Security Administration of the U.S. Department of Energy at Los
Alamos National Laboratory under Contract No. DE-AC52-06NA25396.

{\small
\bibliographystyle{unsrt}
\bibliography{SmartGrid}
}

\end{document}